\documentclass[a4paper,fleqn,usenatbib]{mnras}
\usepackage{newtxtext,newtxmath}
\usepackage[T1]{fontenc}
\usepackage{ae,aecompl}
\usepackage{graphicx}	
\usepackage{amsmath}	

\usepackage{amssymb}	

\title[Frequency-dependent damping of coronal slow waves]{A new look at the frequency-dependent damping of slow-mode waves in the solar corona}
\author[D.~Y. Kolotkov \& V.~M. Nakariakov]{
Dmitrii Y. Kolotkov$^{1}$\thanks{E-mail: D.Kolotkov.1@warwick.ac.uk (DYK)},
Valery M. Nakariakov$^{1}$,
\\
$^{1}$Centre for Fusion, Space and Astrophysics, Physics Department, University of Warwick, Coventry CV4 7AL, United Kingdom
}

\date{Accepted XXX. Received YYY; in original form ZZZ}

\pubyear{2022}

\begin{document}
\label{firstpage}
\pagerange{\pageref{firstpage}--\pageref{lastpage}}
\maketitle

\begin{abstract}
Being directly observed in the Doppler shift and imaging data and indirectly as quasi-periodic pulsations in solar and stellar flares, slow magnetoacoustic waves offer an important seismological tool for probing many vital parameters of the coronal plasma. A recently understood active nature of the solar corona for magnetoacoustic waves, manifested through the phenomenon of wave-induced thermal misbalance, led to the identification of new natural mechanisms for the interpretation of observed properties of waves. A frequency-dependent damping of slow waves in various coronal plasma structures remains an open question, as traditional wave damping theories fail to match observations. We demonstrate that accounting for the back-reaction caused by thermal misbalance on the wave dynamics leads to a modification of the relationship between the damping time and oscillation period of standing slow waves, prescribed by the linear theory. The modified relationship is not of a power-law form and has the equilibrium plasma conditions and properties of the coronal heating/cooling processes as free parameters. It is shown to readily explain the observed scaling of the damping time with period of standing slow waves in hot coronal loops. Functional forms of the unknown coronal heating process, consistent with the observed frequency-dependent damping, are seismologically revealed.
\end{abstract}




\section{Introduction} \label{sec:intro}
Slow magnetoacoustic waves are ubiquitously present in various plasma structures of the solar atmosphere and provide us with an important tool for coronal seismology \citep[see e.g.][for comprehensive reviews]{2012RSPTA.370.3193D, 2020ARA&A..58..441N}. These waves are observed to have relatively long oscillation periods (from several minutes to several tens of minutes) and comparable damping times, are highly sensitive to local thermodynamic and magnetohydrodynamic (MHD) parameters of the background plasma, and are known to appear in propagating, standing, or sloshing forms \citep{2009SSRv..149...65D, 2011SSRv..158..267B, 2011SSRv..158..397W, 2016GMS...216..395W, 2016GMS...216..419B, 2019ApJ...874L...1N, 2021SSRv..217...34W, 2021SSRv..217...76B}. Observations of slow waves are used for seismological probing of the effective polytropic index \citep{2011ApJ...727L..32V, 2018ApJ...868..149K} and coefficients of such transport processes as thermal conduction and viscosity \citep{2015ApJ...811L..13W, 2018ApJ...860..107W} in the corona. In hot flaring loops, slow waves can be used for diagnostics of the coronal magnetic field \citep{2007ApJ...656..598W, 2015A&A...581A..78Z, 2016NatPh..12..179J, 2017A&A...600A..37N}. Also, the internal thermal structuring of coronal loops can be inferred seismologically by slow waves \citep[e.g.][]{2003A&A...404L...1K, 2017ApJ...834..103K}. A strong link between the wave propagation direction and the local direction of the magnetic field can be used for revealing the latter by observations of slow waves \citep{2009ApJ...697.1674M}. This possibility seems to be especially important for obtaining a 3D structure of the coronal magnetic field in the absence of stereoscopic observations.
In addition, the periodic or quasi-periodic nature of slow waves, caused either by the effect of dispersion \citep{2019PhPl...26h2113Z} or as a standing mode in an MHD resonator \citep[e.g.][]{2002ApJ...580L..85O, 2015ApJ...807...98Y}, can lead to a modulation of the electromagnetic emission from solar (and stellar) flares in the form of quasi-periodic pulsations (see e.g. \citealt{2006A&A...446.1151N}, and a series of recent comprehensive reviews by \citealt{2016SoPh..291.3143V, 2018SSRv..214...45M, 2020STP.....6a...3K, 2021SSRv..217...66Z}).

Another important application of slow waves in the corona is connected with their potential to probe the enigmatic coronal heating and cooling processes \citep{2020SSRv..216..140V, 2021PPCF...63l4008K}. Indeed, being essentially compressive, slow waves perturb not only the mechanical equilibrium of the coronal plasma, but also its delicate local thermal balance. The back-reaction that the wave experiences from a perturbed thermal equilibrium, also known as \emph{thermal misbalance}, was shown to have strong impact on the wave dynamics thus making the corona an active medium for magnetoacoustic waves. In particular, the phenomenon of wave-induced thermal misbalance was previously shown to cause an enhanced or suppressed damping of slow waves, or even their amplification, depending on the parameters of the coronal heating/cooling model \citep{2017ApJ...849...62N, 2019A&A...628A.133K, 2021A&A...646A.155D}. In all these works, damping of the lowest slow wave harmonic was addressed only, without taking its frequency-dependent nature into account.
\citet{2019PhPl...26h2113Z, 2021SoPh..296...96Z} demonstrated that the slow wave damping caused by thermal misbalance is accompanied by the phenomenon of wave dispersion and modifications in the characteristic sound and tube speeds. \citet{2021SoPh..296..122B} showed that the dispersion of slow waves, caused by thermal misbalance, may be much more efficient than the geometric waveguide-caused dispersion, especially at longer periods. \citet{2021SoPh..296..105P, 2022SoPh..297....5P} studied the role of thermal misbalance in the phase behaviour of slow waves. Considering the stability of slow and thermal (entropy) modes as a necessary condition for the existence of a long-lived hot corona, \citet{2020A&A...644A..33K} obtained seismological constraints on the functional form of a steady coronal heating. Likewise, observations of quasi-periodic pulsations in hot transient loop systems and their modelling in terms of slow waves allowed for constraining the location and duration of impulsive heating events \citep{2019ApJ...884..131R}.

One of the outstanding puzzles about slow waves in the solar corona concerns the nature of their frequency-dependent damping. For example, observations of standing \citep[e.g.][]{2008ApJ...685.1286V, 2011SSRv..158..397W, 2019ApJ...874L...1N} and propagating \citep[e.g.][]{2014ApJ...789..118K, 2016GMS...216..419B, 2021SSRv..217...76B} slow waves in coronal loops demonstrate an almost linear scaling between the damping time/length and oscillation period (shorter-period waves decay faster). Moreover, a statistically similar linear proportionality was detected for the damping time and oscillation period of rapidly decaying long-period quasi-periodic pulsations, observed in the decay phase of solar and stellar flares and associated with slow waves \citep{2016ApJ...830..110C}.
While at periods shorter than 10\,min the observed linear scaling of the damping time $\tau_\mathrm{D}$ with the period $P$ in loops could be explained by the change of the regime of field-aligned thermal conduction from \lq\lq higher\rq\rq\ (in which $\tau_\mathrm{D}\propto P^0$) to \lq\lq lower\rq\rq\ (in which $\tau_\mathrm{D}\propto P^2$), see e.g. Fig.~16 in \citet{2021SSRv..217...34W} and also \citet{2016ApJ...820...13M}, the linear scaling seen at longer periods cannot be explained by the standard theories of thermal conduction and/or viscosity both of which predict $\tau_\mathrm{D}\propto P^2$. Likewise, the frequency-dependent damping of propagating slow waves in polar regions, for which both positive \citep{2014A&A...568A..96G} and negative \citep{2014ApJ...789..118K, 2018ApJ...853..134M} correlations between damping length and period were observed, remains another open question.

In this Letter, we present a paradigm changing result that the observed frequency-dependent damping of slow magnetoacoustic waves in the solar corona does not necessarily obey a power-law dependence prescribed by traditional theories of thermal conduction or viscosity, for example, and the use of the latter for the interpretation of observations may thus be misleading. We show that accounting for the back-reaction of wave-caused perturbation of the local thermal equilibrium, as an intrinsic property of the coronal plasma, modifies the theoretical relationship between the damping time and oscillation period of slow waves. The modified relationship is shown to be well consistent with observations of standing slow waves in coronal loops. Moreover, we demonstrate that the frequency-dependent damping of slow waves carries an important information about thermodynamic properties of the coronal plasma, and could be used as a new proxy of the enigmatic coronal heating function.

\section{Field-aligned thermal conduction \& wave-induced thermal misbalance} \label{sec:theory}

Slow magnetoacoustic waves in solar coronal loops are well known to be highly sensitive to the effects of field-aligned thermal conduction, which is traditionally considered as the major cause of wave damping (see e.g. \citealt{2003A&A...408..755D}, and the most recent reviews by \citealt{2021SSRv..217...34W, 2021SSRv..217...76B}).
In more recent works \citep[e.g.][]{2019A&A...628A.133K, 2021A&A...646A.155D}, the phenomenon of wave-induced misbalance between optically thin radiation and some unspecified external heating was shown to be capable of causing damping of slow waves comparable to that caused by thermal conduction or even stronger.
Thus, the dynamics of linear slow waves in coronal plasma structures, coupled to the entropy (thermal) mode through the non-adiabatic effects such as thermal conduction and thermal misbalance, can be described by a polynomial dispersion relation for the cyclic frequency $\omega$ and the wavenumber $k$. This dispersion relation has been derived and discussed in detail in a series of previous works \citep[see e.g.][for the most recent papers]{2022SoPh..297....5P, 2021SoPh..296..122B, 2019PhPl...26h2113Z}, hence we do not present its full form and derivation in this work. In the limit of weak (\lq\lq lower\rq\rq) non-adiabaticity, in which the wave is only mildly affected by non-adiabatic processes, and for the plasma parameter $\beta \to 0$ (infinite field approximation) typical for coronal conditions, the dispersion relation for standing slow waves with real $k$ and complex $\omega = \omega_\mathrm{R} + i\omega_\mathrm{I}$ reduces to \citep[e.g.][]{2019A&A...628A.133K, 2021A&A...646A.155D},
\begin{align}
	&\omega_\mathrm{R} = c_\mathrm{s}k,\label{eq:omega_r}\\
	&\omega_\mathrm{I} = \frac{1}{2}\left(\frac{\gamma-1}{\gamma}\frac{1}{\tau_\mathrm{cond}}+\frac{1}{\tau_\mathrm{M}}\right).\label{eq:omega_i}
\end{align}
In Eqs.~(\ref{eq:omega_r})--(\ref{eq:omega_i}), $c_\mathrm{s}$ is the standard sound speed, $\tau_\mathrm{cond}=\rho_0 C_V k^{-2}/ \kappa_\parallel$ is the characteristic wavelength-dependent timescale of the field-aligned thermal conduction with the coefficient $\kappa_\parallel = 10^{-11}T_0^{5/2}$\,W\,m$^{-1}$\,K$^{-1}$, and $\tau_\mathrm{M} = \gamma C_V/\left[(\gamma-1)Q_T+(\rho_0/T_0)Q_{\rho}\right]$ is the timescale of thermal misbalance, determined by the derivatives of the net heat-loss function $Q$ with respect to the plasma temperature $T$ and density $\rho$, evaluated at the equilibrium; $C_V = (\gamma - 1)^{-1}k_\mathrm{B}/m$, $\gamma$, and $m$ are the specific heat capacity, adiabatic index, and mean particle mass, respectively.

From Eqs.~(\ref{eq:omega_r})--(\ref{eq:omega_i}), one can obtain the relationship between the oscillation period $P=2\pi/\omega_\mathrm{R}$ and damping time $\tau_\mathrm{D}=1/\omega_\mathrm{I}$ of standing slow waves, as
\begin{equation}\label{eq:damp-per}
	\tau_\mathrm{D}=\frac{2\tau_\mathrm{M}P^2}{d\tau_\mathrm{M} + P^2},
\end{equation}
with the coefficient $d=4\pi^2(\gamma-1)\kappa_\parallel/\gamma\rho_0C_Vc_\mathrm{s}^2$ \citep{2003A&A...408..755D}.
In the limit $\tau_\mathrm{M} \gg P$ (and $\gg d$), i.e. when the effect of wave-induced thermal misbalance is practically absent, Eq.~(\ref{eq:damp-per}) reduces to the well-known quadratic proportionality between the oscillation period and damping time, $\tau_\mathrm{D}\propto P^2$, which has been shown to be generally inconsistent with observations of slow waves in the corona \citep[e.g.][]{2019ApJ...874L...1N, 2021SSRv..217...34W, 2021SSRv..217...76B}.
However, in standing slow waves observed in the corona, $d\approx20$\,min, i.e. about $P$. Moreover, \citet{2020A&A...644A..33K} demonstrated that for a broad range of typical coronal conditions, in which slow waves are observed either in a standing or propagating form, $\tau_\mathrm{M}$ is from several minutes to several tens of minutes, i.e. about the wave oscillation period $P$ too. The latter estimations, i.e. the facts that both coefficients $d$ and $\tau_\mathrm{M}$ are about $P$ and, hence, $d\tau_\mathrm{M}$ is about $P^2$, do not allow one to use either $P^2/d\tau_\mathrm{M}$ or its reciprocal as a small parameter in Eq.~(\ref{eq:damp-per}).

\section{Comparison with observations} \label{sec:obs}

\begin{figure}
	\centering
	\includegraphics[width=\linewidth]{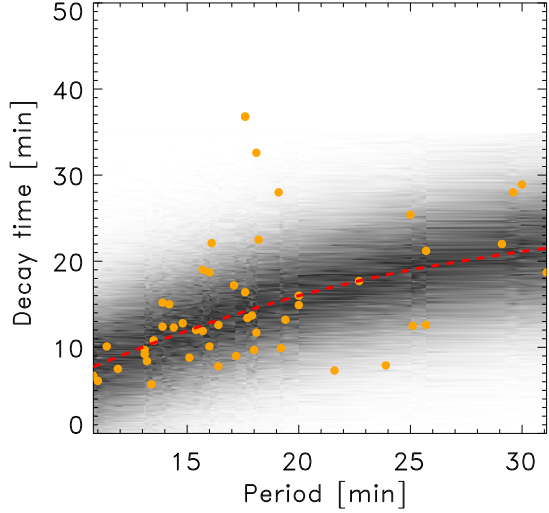}
	\caption{Scaling of the damping time of standing slow magnetoacoustic waves with oscillation period, observed by SOHO/SUMER in coronal loops at 6.3\,MK \citep[the orange circles, see][]{2019ApJ...874L...1N} and best-fitted by the function given by Eq.~(\ref{eq:damp-per}) using MCMC approach (the red dashed line). The grey shading shows the posterior predictive distribution obtained with MCMC, which reflects possible variations of the best-fit model parameters.
	}
	\label{fig:damp-per}
\end{figure}

For the validation of Eq.~(\ref{eq:damp-per}), we use observations of standing slow waves in coronal loops, previously reported in the research literature and summarised in \citet{2019ApJ...874L...1N}. More specifically, rapidly decaying standing slow waves have been observed in the Doppler shift (e.g. with SOHO/SUMER, Yohkoh/BCS, Hinode/EIS) and imaging (e.g. with SDO/AIA) data, at temperatures ranging from 2\,MK to 14\,MK, with most of the events observed by SUMER at 6.3\,MK. As the relationship described by Eq.~(\ref{eq:damp-per}) seems to be highly sensitive to the plasma temperature, in this work we focus on the events observed at 6.3\,MK only, as the most frequent detections. Thus, the statistics of damping times and oscillation periods (and apparent dependence between those statistics) of standing slow waves obtained from SUMER observations of hot coronal loops at 6.3\,MK are shown in Fig.~\ref{fig:damp-per}.

We fit the observed scaling of the damping time with the oscillation period by the model prescribed by Eq.~(\ref{eq:damp-per}), fixing the plasma temperature $T_0$ to 6.3\,MK and treating the plasma density $\rho_0$ and characteristic time of thermal misbalance $\tau_\mathrm{M}$ as free parameters. For fitting, we used Bayesian inference with Markov chain Monte Carlo (MCMC) sampling approach, implemented by the Solar Bayesian Analysis Toolkit \citep[][]{2021ApJS..252...11A}. The best-fit curve is obtained with MCMC for $\rho_0 = 4.0^{+4.1}_{-1.1}\times10^{-12}$\,kg\,m$^{-3}$ and $\tau_\mathrm{M}=14.2^{+4.9}_{-4.5}$\,min.
The corresponding posterior predictive distribution (i.e. where the data points are expected according to the model) and the best-fitting curve are shown in Fig.~\ref{fig:damp-per} and seem to be in a reasonable agreement with observations.
{The apparent scattering of data points in Fig.~\ref{fig:damp-per} around the best-fit curve can be attributed to variations of the plasma parameters of oscillating loops, which are accounted for by the estimation uncertainties obtained with MCMC as demonstrated by the posterior probability distribution.}
\begin{figure}
	\centering
	\includegraphics[width=\linewidth]{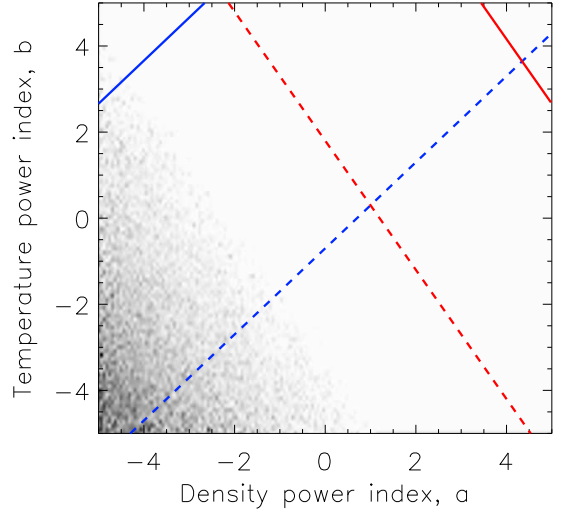}
	\caption{Power indices $a$ and $b$ in the parametrisation of the coronal heating function $\mathcal{H}\propto \rho^aT^b$, for which Eqs.~(\ref{eq:damp-per}) and (\ref{eq:tau_m}) are consistent with the observed frequency-dependent damping, as shown in Fig.~\ref{fig:damp-per} (the grey shading).
	For comparison, the red and blue lines show (essentially broader) intervals of $a$ and $b$, below which the acoustic (red) and thermal (blue) modes remain stable to small-amplitude perturbations, according to Eqs.~(9)--(10) in \citet{2020A&A...644A..33K}, with thermal conduction and the loop length of 300\,Mm (the solid lines) and without thermal conduction (the dashed lines). For longer/shorter loops, the solid lines appear closer to/further from the dashed lines, respectively.
	}
	\label{fig:damp-ab}
\end{figure}

Using the link of the characteristic misbalance time $\tau_\mathrm{M}$ with the heat-loss function in Eqs.~(\ref{eq:omega_i}) and (\ref{eq:damp-per}), we can identify properties of the unknown coronal heating function, which would be consistent with the observed frequency-dependent damping of slow waves. Indeed, considering the coronal heat-loss function $Q$ as a net effect between local energy losses $\mathcal{L}$ by optically thin radiation and energy gains by some unspecified steady heating $\mathcal{H}$, $Q=\mathcal{L}-\mathcal{H}$, and using the atomic database CHIANTI to model $\mathcal{L}$ \citep{1997A&AS..125..149D, 2021ApJ...909...38D}, the unknown heating rate $\mathcal{H}$ could be parametrised by a power-law function of the local plasma conditions, $\mathcal{H} \propto \rho^a T^b$ \citep{1965ApJ...142..531F, 1978ApJ...220..643R, 1988SoPh..117...51D, 1993ApJ...415..335I}. A possible dependence of the heating function on the local value of the magnetic field was recently shown to have no effect on the dynamics of slow waves in the regime of low plasma-$\beta$ \citep{2021A&A...646A.155D}, hence it is omitted. The proportionality coefficient in this parametrisation of $\mathcal{H}$ is obtained from the initial thermal equilibrium condition $Q_0=0$, implying a balance between unperturbed rates of radiative losses and heating. With this, the characteristic time of the wave-induced thermal misbalance $\tau_\mathrm{M}$ can be rewritten as
\begin{equation}\label{eq:tau_m}
	\tau_\mathrm{M}=\frac{\gamma}{\gamma-1}\frac{C_V}{\dfrac{\partial \mathcal{L}_0}{\partial T}-\dfrac{\mathcal{L}_0}{T_0}\left(\dfrac{a-1}{\gamma-1}+b\right)},
\end{equation}
where $\mathcal{L}_0$ is the equilibrium value of the radiative loss rate, modelled by CHIANTI, and the derivative $\partial \mathcal{L}_0/\partial T$ characterises its slope at the equilibrium value of the coronal temperature. Using Eq.~(\ref{eq:tau_m}), we rewrite the distribution of the values of $\tau_\mathrm{M}$ sampled by MCMC (Fig.~\ref{fig:damp-per}) in terms of the heating power indices $a$ and $b$ which are treated as free parameters. Figure~\ref{fig:damp-ab} shows the heating models in the parametric plane $(a,b)$, which are consistent with the observed frequency-dependent damping of standing slow waves in hot coronal loops. This result puts new important constraints onto the solar coronal heating function in comparison with those obtained by \citet{2020A&A...644A..33K}, which were based on the stability of fundamental acoustic and thermal modes only and did not take a frequency-dependent nature of acoustic damping into account.
{In particular, a specific regime of thermal misbalance with constant heating \citep[i.e. with $a=0$ and $b=0$, see e.g.][]{2004A&A...415..705D} does not seem capable of explaining the observed frequency-dependent damping.}

\section{Discussion and Conclusions} \label{sec:discussion}

In this Letter, we presented a new look at the phenomenon of frequency-dependent damping of slow magnetoacoustic waves in the solar corona through the prism of recently understood effects of wave-induced thermal misbalance. In contrast to traditional approaches used to interpret the observed scaling of the damping time $\tau_\mathrm{D}$ of slow oscillations with the period $P$ by a power-law dependence (e.g. $\tau_\mathrm{D}\propto P^2$ in the standard theory of wave damping by field-aligned thermal conduction), we demonstrated that the thermodynamically active nature of the solar corona and effects of thermal misbalance modify this dependence to $\tau_\mathrm{D}\propto P^2/(1+P^2)$, with the coefficients determined by the equilibrium parameters of the plasma and properties of coronal heating and cooling processes. A quantitative comparison of this new relationship with observations showed a good agreement for a reasonable choice of the model parameters, which allowed for refining the existing seismological constraints on the coronal heating function. More specifically, our findings can be summarised as follows.
\begin{itemize}
	\item Since the plasma of the corona of the Sun acts as an active medium for magnetoacoustic waves (due to a continuous interplay between the processes of coronal heating and cooling, sustaining a long-lived corona), the description and interpretation of the observed frequency-dependent damping of slow magnetoacoustic waves by power-law models predicted by e.g. field-aligned thermal conduction may be incomplete or even inconsistent. Accounting for the effects of wave-induced thermal misbalance in the theoretical treatment of a frequency-dependent damping of linear slow waves allowed us to obtain a new relationship between the damping time and period, given by Eq.~(\ref{eq:damp-per}). This new relationship is not of a power-law form, and has the equilibrium plasma density $\rho_0$, temperature $T_0$, and characteristic time of thermal misbalance $\tau_\mathrm{M}$ as free parameters.
	
	\item Using MCMC approach, the modified relationship was shown to readily explain the scaling between the damping time of slow oscillations and period, observed by SOHO/SUMER in hot coronal loops ($T_0=6.3$\,MK), for $\rho_0 = 4.0^{+4.1}_{-1.1}\times10^{-12}$\,kg\,m$^{-3}$ and $\tau_\mathrm{M}=14.2^{+4.9}_{-4.5}$\,min (see Fig.~\ref{fig:damp-per}). The seismologically revealed mean value of the plasma density $\rho_0$ is well consistent with previous estimations of oscillating loops' density, performed with independent methods \citep[e.g. through the differential emission measure analysis, see][]{2015ApJ...811L..13W}. Likewise, the obtained value of $\tau_\mathrm{M}$ is about the period of slow oscillations typically observed in the corona, which justifies the need for taking the effect of thermal misbalance into account in Eq.~(\ref{eq:damp-per}).
	
	\item In contrast to \citet{2020A&A...644A..33K}, who calculated $\tau_\mathrm{M}$ directly assuming a specific form of radiative function and guessed heating model, seismological estimation of $\tau_\mathrm{M}$ presented in this work is based on observations, does not require any a priori assumptions on the coronal heating/cooling model, and is therefore more generic. If we, however, assume the CHIANTI model for optically thin radiative losses and present the unknown heating function in the power law form, $\mathcal{H}\propto \rho^a T^b$, the distribution of $\tau_\mathrm{M}$ sampled by MCMC in our seismological analysis allows for identifying the heating parameters $a$ and $b$ which would allow for the observed frequency-dependent damping of slow waves (see Fig.~\ref{fig:damp-ab}).
	
	\item The performed analysis is focussed on standing slow waves in hot coronal loops. However, propagating slow magnetoacoustic waves in quiescent loops, polar plumes, and interplume regions are also observed to manifest a frequency-dependent damping \citep[see e.g.][]{2014ApJ...789..118K, 2018ApJ...853..134M}, for which a relationship analogous to Eq.~(\ref{eq:damp-per}) could be derived and used for the interpretation. Of a particular interest is the role of a newly revealed form of frequency-dependent damping in the formation of sloshing slow magnetoacoustic oscillations \citep[e.g.][]{2013ApJ...779L...7K, 2015ApJ...804....4K, 2016ApJ...826L..20R, 2019ApJ...874L...1N}. Likewise, the presented study could be generalised for non-zero values of the plasma parameter $\beta$, effects of moderate or strong non-adiabaticity, and nonlinear regimes of wave damping. 

\end{itemize}

\section*{Acknowledgements}
D.Y.K. and V.M.N. acknowledge support from the STFC consolidated grant ST/T000252/1.

\section*{Data availability}
The data underlying this article are available in the article and in the references therein.

\bibliographystyle{mnras}

\begin{thebibliography}{}
	\makeatletter
	\relax
	\def\mn@urlcharsother{\let\do\@makeother \do\$\do\&\do\#\do\^\do\_\do\%\do\~}
	\def\mn@doi{\begingroup\mn@urlcharsother \@ifnextchar [ {\mn@doi@}
		{\mn@doi@[]}}
	\def\mn@doi@[#1]#2{\def\@tempa{#1}\ifx\@tempa\@empty \href
		{http://dx.doi.org/#2} {doi:#2}\else \href {http://dx.doi.org/#2} {#1}\fi
		\endgroup}
	\def\mn@eprint#1#2{\mn@eprint@#1:#2::\@nil}
	\def\mn@eprint@arXiv#1{\href {http://arxiv.org/abs/#1} {{\tt arXiv:#1}}}
	\def\mn@eprint@dblp#1{\href {http://dblp.uni-trier.de/rec/bibtex/#1.xml}
		{dblp:#1}}
	\def\mn@eprint@#1:#2:#3:#4\@nil{\def\@tempa {#1}\def\@tempb {#2}\def\@tempc
		{#3}\ifx \@tempc \@empty \let \@tempc \@tempb \let \@tempb \@tempa \fi \ifx
		\@tempb \@empty \def\@tempb {arXiv}\fi \@ifundefined
		{mn@eprint@\@tempb}{\@tempb:\@tempc}{\expandafter \expandafter \csname
			mn@eprint@\@tempb\endcsname \expandafter{\@tempc}}}
	
	\bibitem[\protect\citeauthoryear{{Anfinogentov}, {Nakariakov}, {Pascoe}  \&
		{Goddard}}{{Anfinogentov} et~al.}{2021}]{2021ApJS..252...11A}
	{Anfinogentov} S.~A.,  {Nakariakov} V.~M.,  {Pascoe} D.~J.,   {Goddard} C.~R.,
	2021, \mn@doi [\apjs] {10.3847/1538-4365/abc5c1}, \href
	{https://ui.adsabs.harvard.edu/abs/2021ApJS..252...11A} {252, 11}
	
	\bibitem[\protect\citeauthoryear{{Banerjee} \& {Krishna Prasad}}{{Banerjee} \&
		{Krishna Prasad}}{2016}]{2016GMS...216..419B}
	{Banerjee} D.,  {Krishna Prasad} S.,  2016, \mn@doi [Washington DC American
	Geophysical Union Geophysical Monograph Series] {10.1002/9781119055006.ch24},
	\href {https://ui.adsabs.harvard.edu/abs/2016GMS...216..419B} {216, 419}
	
	\bibitem[\protect\citeauthoryear{{Banerjee}, {Gupta}  \& {Teriaca}}{{Banerjee}
		et~al.}{2011}]{2011SSRv..158..267B}
	{Banerjee} D.,  {Gupta} G.~R.,   {Teriaca} L.,  2011, \mn@doi [\ssr]
	{10.1007/s11214-010-9698-z}, \href
	{https://ui.adsabs.harvard.edu/abs/2011SSRv..158..267B} {158, 267}
	
	\bibitem[\protect\citeauthoryear{{Banerjee} et~al.,}{{Banerjee}
		et~al.}{2021}]{2021SSRv..217...76B}
	{Banerjee} D.,  et~al., 2021, \mn@doi [\ssr] {10.1007/s11214-021-00849-0},
	\href {https://ui.adsabs.harvard.edu/abs/2021SSRv..217...76B} {217, 76}
	
	\bibitem[\protect\citeauthoryear{{Belov}, {Molevich}  \&
		{Zavershinskii}}{{Belov} et~al.}{2021}]{2021SoPh..296..122B}
	{Belov} S.~A.,  {Molevich} N.~E.,   {Zavershinskii} D.~I.,  2021, \mn@doi
	[\solphys] {10.1007/s11207-021-01868-4}, \href
	{https://ui.adsabs.harvard.edu/abs/2021SoPh..296..122B} {296, 122}
	
	\bibitem[\protect\citeauthoryear{{Cho}, {Cho}, {Nakariakov}, {Kim}  \&
		{Kumar}}{{Cho} et~al.}{2016}]{2016ApJ...830..110C}
	{Cho} I.~H.,  {Cho} K.~S.,  {Nakariakov} V.~M.,  {Kim} S.,   {Kumar} P.,  2016,
	\mn@doi [\apj] {10.3847/0004-637X/830/2/110}, \href
	{https://ui.adsabs.harvard.edu/abs/2016ApJ...830..110C} {830, 110}
	
	\bibitem[\protect\citeauthoryear{{Dahlburg} \& {Mariska}}{{Dahlburg} \&
		{Mariska}}{1988}]{1988SoPh..117...51D}
	{Dahlburg} R.~B.,  {Mariska} J.~T.,  1988, \mn@doi [\solphys]
	{10.1007/BF00148571}, \href
	{https://ui.adsabs.harvard.edu/abs/1988SoPh..117...51D} {117, 51}
	
	\bibitem[\protect\citeauthoryear{{De Moortel}}{{De
			Moortel}}{2009}]{2009SSRv..149...65D}
	{De Moortel} I.,  2009, \mn@doi [\ssr] {10.1007/s11214-009-9526-5}, \href
	{https://ui.adsabs.harvard.edu/abs/2009SSRv..149...65D} {149, 65}
	
	\bibitem[\protect\citeauthoryear{{De Moortel} \& {Hood}}{{De Moortel} \&
		{Hood}}{2003}]{2003A&A...408..755D}
	{De Moortel} I.,  {Hood} A.~W.,  2003, \mn@doi [\aap]
	{10.1051/0004-6361:20030984}, \href
	{https://ui.adsabs.harvard.edu/abs/2003A&A...408..755D} {408, 755}
	
	\bibitem[\protect\citeauthoryear{{De Moortel} \& {Hood}}{{De Moortel} \&
		{Hood}}{2004}]{2004A&A...415..705D}
	{De Moortel} I.,  {Hood} A.~W.,  2004, \mn@doi [\aap]
	{10.1051/0004-6361:20034233}, \href
	{https://ui.adsabs.harvard.edu/abs/2004A&A...415..705D} {415, 705}
	
	\bibitem[\protect\citeauthoryear{{De Moortel} \& {Nakariakov}}{{De Moortel} \&
		{Nakariakov}}{2012}]{2012RSPTA.370.3193D}
	{De Moortel} I.,  {Nakariakov} V.~M.,  2012, \mn@doi [Philosophical
	Transactions of the Royal Society of London Series A]
	{10.1098/rsta.2011.0640}, \href
	{https://ui.adsabs.harvard.edu/abs/2012RSPTA.370.3193D} {370, 3193}
	
	\bibitem[\protect\citeauthoryear{{Del Zanna}, {Dere}, {Young}  \& {Landi}}{{Del
			Zanna} et~al.}{2021}]{2021ApJ...909...38D}
	{Del Zanna} G.,  {Dere} K.~P.,  {Young} P.~R.,   {Landi} E.,  2021, \mn@doi
	[\apj] {10.3847/1538-4357/abd8ce}, \href
	{https://ui.adsabs.harvard.edu/abs/2021ApJ...909...38D} {909, 38}
	
	\bibitem[\protect\citeauthoryear{{Dere}, {Landi}, {Mason}, {Monsignori Fossi}
		\& {Young}}{{Dere} et~al.}{1997}]{1997A&AS..125..149D}
	{Dere} K.~P.,  {Landi} E.,  {Mason} H.~E.,  {Monsignori Fossi} B.~C.,   {Young}
	P.~R.,  1997, \mn@doi [\aaps] {10.1051/aas:1997368}, \href
	{https://ui.adsabs.harvard.edu/abs/1997A&AS..125..149D} {125, 149}
	
	\bibitem[\protect\citeauthoryear{{Duckenfield}, {Kolotkov}  \&
		{Nakariakov}}{{Duckenfield} et~al.}{2021}]{2021A&A...646A.155D}
	{Duckenfield} T.~J.,  {Kolotkov} D.~Y.,   {Nakariakov} V.~M.,  2021, \mn@doi
	[\aap] {10.1051/0004-6361/202039791}, \href
	{https://ui.adsabs.harvard.edu/abs/2021A&A...646A.155D} {646, A155}
	
	\bibitem[\protect\citeauthoryear{{Field}}{{Field}}{1965}]{1965ApJ...142..531F}
	{Field} G.~B.,  1965, \mn@doi [\apj] {10.1086/148317}, \href
	{https://ui.adsabs.harvard.edu/abs/1965ApJ...142..531F} {142, 531}
	
	\bibitem[\protect\citeauthoryear{{Gupta}}{{Gupta}}{2014}]{2014A&A...568A..96G}
	{Gupta} G.~R.,  2014, \mn@doi [\aap] {10.1051/0004-6361/201323200}, \href
	{https://ui.adsabs.harvard.edu/abs/2014A&A...568A..96G} {568, A96}
	
	\bibitem[\protect\citeauthoryear{{Ibanez S.} \& {Escalona T.}}{{Ibanez S.} \&
		{Escalona T.}}{1993}]{1993ApJ...415..335I}
	{Ibanez S.} M.~H.,  {Escalona T.} O.~B.,  1993, \mn@doi [\apj]
	{10.1086/173167}, \href
	{https://ui.adsabs.harvard.edu/abs/1993ApJ...415..335I} {415, 335}
	
	\bibitem[\protect\citeauthoryear{{Jess} et~al.,}{{Jess}
		et~al.}{2016}]{2016NatPh..12..179J}
	{Jess} D.~B.,  et~al., 2016, \mn@doi [Nature Physics] {10.1038/nphys3544},
	\href {https://ui.adsabs.harvard.edu/abs/2016NatPh..12..179J} {12, 179}
	
	\bibitem[\protect\citeauthoryear{{King}, {Nakariakov}, {Deluca}, {Golub}  \&
		{McClements}}{{King} et~al.}{2003}]{2003A&A...404L...1K}
	{King} D.~B.,  {Nakariakov} V.~M.,  {Deluca} E.~E.,  {Golub} L.,   {McClements}
	K.~G.,  2003, \mn@doi [\aap] {10.1051/0004-6361:20030763}, \href
	{https://ui.adsabs.harvard.edu/abs/2003A&A...404L...1K} {404, L1}
	
	\bibitem[\protect\citeauthoryear{{Kolotkov}, {Nakariakov}  \&
		{Zavershinskii}}{{Kolotkov} et~al.}{2019}]{2019A&A...628A.133K}
	{Kolotkov} D.~Y.,  {Nakariakov} V.~M.,   {Zavershinskii} D.~I.,  2019, \mn@doi
	[\aap] {10.1051/0004-6361/201936072}, \href
	{https://ui.adsabs.harvard.edu/abs/2019A&A...628A.133K} {628, A133}
	
	\bibitem[\protect\citeauthoryear{{Kolotkov}, {Duckenfield}  \&
		{Nakariakov}}{{Kolotkov} et~al.}{2020}]{2020A&A...644A..33K}
	{Kolotkov} D.~Y.,  {Duckenfield} T.~J.,   {Nakariakov} V.~M.,  2020, \mn@doi
	[\aap] {10.1051/0004-6361/202039095}, \href
	{https://ui.adsabs.harvard.edu/abs/2020A&A...644A..33K} {644, A33}
	
	\bibitem[\protect\citeauthoryear{{Kolotkov}, {Zavershinskii}  \&
		{Nakariakov}}{{Kolotkov} et~al.}{2021}]{2021PPCF...63l4008K}
	{Kolotkov} D.~Y.,  {Zavershinskii} D.~I.,   {Nakariakov} V.~M.,  2021, \mn@doi
	[Plasma Physics and Controlled Fusion] {10.1088/1361-6587/ac36a5}, \href
	{https://ui.adsabs.harvard.edu/abs/2021PPCF...63l4008K} {63, 124008}
	
	\bibitem[\protect\citeauthoryear{{Krishna Prasad}, {Banerjee}  \& {Van
			Doorsselaere}}{{Krishna Prasad} et~al.}{2014}]{2014ApJ...789..118K}
	{Krishna Prasad} S.,  {Banerjee} D.,   {Van Doorsselaere} T.,  2014, \mn@doi
	[\apj] {10.1088/0004-637X/789/2/118}, \href
	{https://ui.adsabs.harvard.edu/abs/2014ApJ...789..118K} {789, 118}
	
	\bibitem[\protect\citeauthoryear{{Krishna Prasad}, {Jess}, {Klimchuk}  \&
		{Banerjee}}{{Krishna Prasad} et~al.}{2017}]{2017ApJ...834..103K}
	{Krishna Prasad} S.,  {Jess} D.~B.,  {Klimchuk} J.~A.,   {Banerjee} D.,  2017,
	\mn@doi [\apj] {10.3847/1538-4357/834/2/103}, \href
	{https://ui.adsabs.harvard.edu/abs/2017ApJ...834..103K} {834, 103}
	
	\bibitem[\protect\citeauthoryear{{Krishna Prasad}, {Raes}, {Van Doorsselaere},
		{Magyar}  \& {Jess}}{{Krishna Prasad} et~al.}{2018}]{2018ApJ...868..149K}
	{Krishna Prasad} S.,  {Raes} J.~O.,  {Van Doorsselaere} T.,  {Magyar} N.,
	{Jess} D.~B.,  2018, \mn@doi [\apj] {10.3847/1538-4357/aae9f5}, \href
	{https://ui.adsabs.harvard.edu/abs/2018ApJ...868..149K} {868, 149}
	
	\bibitem[\protect\citeauthoryear{{Kumar}, {Innes}  \& {Inhester}}{{Kumar}
		et~al.}{2013}]{2013ApJ...779L...7K}
	{Kumar} P.,  {Innes} D.~E.,   {Inhester} B.,  2013, \mn@doi [\apjl]
	{10.1088/2041-8205/779/1/L7}, \href
	{https://ui.adsabs.harvard.edu/abs/2013ApJ...779L...7K} {779, L7}
	
	\bibitem[\protect\citeauthoryear{{Kumar}, {Nakariakov}  \& {Cho}}{{Kumar}
		et~al.}{2015}]{2015ApJ...804....4K}
	{Kumar} P.,  {Nakariakov} V.~M.,   {Cho} K.-S.,  2015, \mn@doi [\apj]
	{10.1088/0004-637X/804/1/4}, \href
	{https://ui.adsabs.harvard.edu/abs/2015ApJ...804....4K} {804, 4}
	
	\bibitem[\protect\citeauthoryear{{Kupriyanova}, {Kolotkov}, {Nakariakov}  \&
		{Kaufman}}{{Kupriyanova} et~al.}{2020}]{2020STP.....6a...3K}
	{Kupriyanova} E.,  {Kolotkov} D.,  {Nakariakov} V.,   {Kaufman} A.,  2020,
	\mn@doi [Solar-Terrestrial Physics] {10.12737/stp-61202001}, \href
	{https://ui.adsabs.harvard.edu/abs/2020STP.....6a...3K} {6, 3}
	
	\bibitem[\protect\citeauthoryear{{Mandal}, {Magyar}, {Yuan}, {Van Doorsselaere}
		\& {Banerjee}}{{Mandal} et~al.}{2016}]{2016ApJ...820...13M}
	{Mandal} S.,  {Magyar} N.,  {Yuan} D.,  {Van Doorsselaere} T.,   {Banerjee} D.,
	2016, \mn@doi [\apj] {10.3847/0004-637X/820/1/13}, \href
	{https://ui.adsabs.harvard.edu/abs/2016ApJ...820...13M} {820, 13}
	
	\bibitem[\protect\citeauthoryear{{Mandal}, {Krishna Prasad}  \&
		{Banerjee}}{{Mandal} et~al.}{2018}]{2018ApJ...853..134M}
	{Mandal} S.,  {Krishna Prasad} S.,   {Banerjee} D.,  2018, \mn@doi [\apj]
	{10.3847/1538-4357/aaa1a3}, \href
	{https://ui.adsabs.harvard.edu/abs/2018ApJ...853..134M} {853, 134}
	
	\bibitem[\protect\citeauthoryear{{Marsh}, {Walsh}  \& {Plunkett}}{{Marsh}
		et~al.}{2009}]{2009ApJ...697.1674M}
	{Marsh} M.~S.,  {Walsh} R.~W.,   {Plunkett} S.,  2009, \mn@doi [\apj]
	{10.1088/0004-637X/697/2/1674}, \href
	{https://ui.adsabs.harvard.edu/abs/2009ApJ...697.1674M} {697, 1674}
	
	\bibitem[\protect\citeauthoryear{{McLaughlin}, {Nakariakov}, {Dominique},
		{Jel{\'\i}nek}  \& {Takasao}}{{McLaughlin}
		et~al.}{2018}]{2018SSRv..214...45M}
	{McLaughlin} J.~A.,  {Nakariakov} V.~M.,  {Dominique} M.,  {Jel{\'\i}nek} P.,
	{Takasao} S.,  2018, \mn@doi [\ssr] {10.1007/s11214-018-0478-5}, \href
	{https://ui.adsabs.harvard.edu/abs/2018SSRv..214...45M} {214, 45}
	
	\bibitem[\protect\citeauthoryear{{Nakariakov} \& {Kolotkov}}{{Nakariakov} \&
		{Kolotkov}}{2020}]{2020ARA&A..58..441N}
	{Nakariakov} V.~M.,  {Kolotkov} D.~Y.,  2020, \mn@doi [\araa]
	{10.1146/annurev-astro-032320-042940}, \href
	{https://ui.adsabs.harvard.edu/abs/2020ARA&A..58..441N} {58, 441}
	
	\bibitem[\protect\citeauthoryear{{Nakariakov} \& {Melnikov}}{{Nakariakov} \&
		{Melnikov}}{2006}]{2006A&A...446.1151N}
	{Nakariakov} V.~M.,  {Melnikov} V.~F.,  2006, \mn@doi [\aap]
	{10.1051/0004-6361:20053944}, \href
	{https://ui.adsabs.harvard.edu/abs/2006A&A...446.1151N} {446, 1151}
	
	\bibitem[\protect\citeauthoryear{{Nakariakov}, {Afanasyev}, {Kumar}  \&
		{Moon}}{{Nakariakov} et~al.}{2017}]{2017ApJ...849...62N}
	{Nakariakov} V.~M.,  {Afanasyev} A.~N.,  {Kumar} S.,   {Moon} Y.~J.,  2017,
	\mn@doi [\apj] {10.3847/1538-4357/aa8ea3}, \href
	{https://ui.adsabs.harvard.edu/abs/2017ApJ...849...62N} {849, 62}
	
	\bibitem[\protect\citeauthoryear{{Nakariakov}, {Kosak}, {Kolotkov},
		{Anfinogentov}, {Kumar}  \& {Moon}}{{Nakariakov}
		et~al.}{2019}]{2019ApJ...874L...1N}
	{Nakariakov} V.~M.,  {Kosak} M.~K.,  {Kolotkov} D.~Y.,  {Anfinogentov} S.~A.,
	{Kumar} P.,   {Moon} Y.~J.,  2019, \mn@doi [\apjl]
	{10.3847/2041-8213/ab0c9f}, \href
	{https://ui.adsabs.harvard.edu/abs/2019ApJ...874L...1N} {874, L1}
	
	\bibitem[\protect\citeauthoryear{{Nistic{\`o}}, {Polito}, {Nakariakov}  \& {Del
			Zanna}}{{Nistic{\`o}} et~al.}{2017}]{2017A&A...600A..37N}
	{Nistic{\`o}} G.,  {Polito} V.,  {Nakariakov} V.~M.,   {Del Zanna} G.,  2017,
	\mn@doi [\aap] {10.1051/0004-6361/201629324}, \href
	{https://ui.adsabs.harvard.edu/abs/2017A&A...600A..37N} {600, A37}
	
	\bibitem[\protect\citeauthoryear{{Ofman} \& {Wang}}{{Ofman} \&
		{Wang}}{2002}]{2002ApJ...580L..85O}
	{Ofman} L.,  {Wang} T.,  2002, \mn@doi [\apjl] {10.1086/345548}, \href
	{https://ui.adsabs.harvard.edu/abs/2002ApJ...580L..85O} {580, L85}
	
	\bibitem[\protect\citeauthoryear{{Prasad}, {Srivastava}  \& {Wang}}{{Prasad}
		et~al.}{2021}]{2021SoPh..296..105P}
	{Prasad} A.,  {Srivastava} A.~K.,   {Wang} T.,  2021, \mn@doi [\solphys]
	{10.1007/s11207-021-01846-w}, \href
	{https://ui.adsabs.harvard.edu/abs/2021SoPh..296..105P} {296, 105}
	
	\bibitem[\protect\citeauthoryear{{Prasad}, {Srivastava}, {Wang}  \&
		{Sangal}}{{Prasad} et~al.}{2022}]{2022SoPh..297....5P}
	{Prasad} A.,  {Srivastava} A.~K.,  {Wang} T.,   {Sangal} K.,  2022, \mn@doi
	[\solphys] {10.1007/s11207-021-01940-z}, \href
	{https://ui.adsabs.harvard.edu/abs/2022SoPh..297....5P} {297, 5}
	
	\bibitem[\protect\citeauthoryear{{Reale}}{{Reale}}{2016}]{2016ApJ...826L..20R}
	{Reale} F.,  2016, \mn@doi [\apjl] {10.3847/2041-8205/826/2/L20}, \href
	{https://ui.adsabs.harvard.edu/abs/2016ApJ...826L..20R} {826, L20}
	
	\bibitem[\protect\citeauthoryear{{Reale}, {Testa}, {Petralia}  \&
		{Kolotkov}}{{Reale} et~al.}{2019}]{2019ApJ...884..131R}
	{Reale} F.,  {Testa} P.,  {Petralia} A.,   {Kolotkov} D.~Y.,  2019, \mn@doi
	[\apj] {10.3847/1538-4357/ab4270}, \href
	{https://ui.adsabs.harvard.edu/abs/2019ApJ...884..131R} {884, 131}
	
	\bibitem[\protect\citeauthoryear{{Rosner}, {Tucker}  \& {Vaiana}}{{Rosner}
		et~al.}{1978}]{1978ApJ...220..643R}
	{Rosner} R.,  {Tucker} W.~H.,   {Vaiana} G.~S.,  1978, \mn@doi [\apj]
	{10.1086/155949}, \href
	{https://ui.adsabs.harvard.edu/abs/1978ApJ...220..643R} {220, 643}
	
	\bibitem[\protect\citeauthoryear{{Van Doorsselaere}, {Wardle}, {Del Zanna},
		{Jansari}, {Verwichte}  \& {Nakariakov}}{{Van Doorsselaere}
		et~al.}{2011}]{2011ApJ...727L..32V}
	{Van Doorsselaere} T.,  {Wardle} N.,  {Del Zanna} G.,  {Jansari} K.,
	{Verwichte} E.,   {Nakariakov} V.~M.,  2011, \mn@doi [\apjl]
	{10.1088/2041-8205/727/2/L32}, \href
	{https://ui.adsabs.harvard.edu/abs/2011ApJ...727L..32V} {727, L32}
	
	\bibitem[\protect\citeauthoryear{{Van Doorsselaere}, {Kupriyanova}  \&
		{Yuan}}{{Van Doorsselaere} et~al.}{2016}]{2016SoPh..291.3143V}
	{Van Doorsselaere} T.,  {Kupriyanova} E.~G.,   {Yuan} D.,  2016, \mn@doi
	[\solphys] {10.1007/s11207-016-0977-z}, \href
	{https://ui.adsabs.harvard.edu/abs/2016SoPh..291.3143V} {291, 3143}
	
	\bibitem[\protect\citeauthoryear{{Van Doorsselaere} et~al.,}{{Van Doorsselaere}
		et~al.}{2020}]{2020SSRv..216..140V}
	{Van Doorsselaere} T.,  et~al., 2020, \mn@doi [\ssr]
	{10.1007/s11214-020-00770-y}, \href
	{https://ui.adsabs.harvard.edu/abs/2020SSRv..216..140V} {216, 140}
	
	\bibitem[\protect\citeauthoryear{{Verwichte}, {Haynes}, {Arber}  \&
		{Brady}}{{Verwichte} et~al.}{2008}]{2008ApJ...685.1286V}
	{Verwichte} E.,  {Haynes} M.,  {Arber} T.~D.,   {Brady} C.~S.,  2008, \mn@doi
	[\apj] {10.1086/591077}, \href
	{https://ui.adsabs.harvard.edu/abs/2008ApJ...685.1286V} {685, 1286}
	
	\bibitem[\protect\citeauthoryear{{Wang}}{{Wang}}{2011}]{2011SSRv..158..397W}
	{Wang} T.,  2011, \mn@doi [\ssr] {10.1007/s11214-010-9716-1}, \href
	{https://ui.adsabs.harvard.edu/abs/2011SSRv..158..397W} {158, 397}
	
	\bibitem[\protect\citeauthoryear{{Wang}}{{Wang}}{2016}]{2016GMS...216..395W}
	{Wang} T.~J.,  2016, \mn@doi [Washington DC American Geophysical Union
	Geophysical Monograph Series] {10.1002/9781119055006.ch23}, \href
	{https://ui.adsabs.harvard.edu/abs/2016GMS...216..395W} {216, 395}
	
	\bibitem[\protect\citeauthoryear{{Wang}, {Innes}  \& {Qiu}}{{Wang}
		et~al.}{2007}]{2007ApJ...656..598W}
	{Wang} T.,  {Innes} D.~E.,   {Qiu} J.,  2007, \mn@doi [\apj] {10.1086/510424},
	\href {https://ui.adsabs.harvard.edu/abs/2007ApJ...656..598W} {656, 598}
	
	\bibitem[\protect\citeauthoryear{{Wang}, {Ofman}, {Sun}, {Provornikova}  \&
		{Davila}}{{Wang} et~al.}{2015}]{2015ApJ...811L..13W}
	{Wang} T.,  {Ofman} L.,  {Sun} X.,  {Provornikova} E.,   {Davila} J.~M.,  2015,
	\mn@doi [\apjl] {10.1088/2041-8205/811/1/L13}, \href
	{https://ui.adsabs.harvard.edu/abs/2015ApJ...811L..13W} {811, L13}
	
	\bibitem[\protect\citeauthoryear{{Wang}, {Ofman}, {Sun}, {Solanki}  \&
		{Davila}}{{Wang} et~al.}{2018}]{2018ApJ...860..107W}
	{Wang} T.,  {Ofman} L.,  {Sun} X.,  {Solanki} S.~K.,   {Davila} J.~M.,  2018,
	\mn@doi [\apj] {10.3847/1538-4357/aac38a}, \href
	{https://ui.adsabs.harvard.edu/abs/2018ApJ...860..107W} {860, 107}
	
	\bibitem[\protect\citeauthoryear{{Wang}, {Ofman}, {Yuan}, {Reale}, {Kolotkov}
		\& {Srivastava}}{{Wang} et~al.}{2021}]{2021SSRv..217...34W}
	{Wang} T.,  {Ofman} L.,  {Yuan} D.,  {Reale} F.,  {Kolotkov} D.~Y.,
	{Srivastava} A.~K.,  2021, \mn@doi [\ssr] {10.1007/s11214-021-00811-0}, \href
	{https://ui.adsabs.harvard.edu/abs/2021SSRv..217...34W} {217, 34}
	
	\bibitem[\protect\citeauthoryear{{Yuan}, {Van Doorsselaere}, {Banerjee}  \&
		{Antolin}}{{Yuan} et~al.}{2015}]{2015ApJ...807...98Y}
	{Yuan} D.,  {Van Doorsselaere} T.,  {Banerjee} D.,   {Antolin} P.,  2015,
	\mn@doi [\apj] {10.1088/0004-637X/807/1/98}, \href
	{https://ui.adsabs.harvard.edu/abs/2015ApJ...807...98Y} {807, 98}
	
	\bibitem[\protect\citeauthoryear{{Zavershinskii}, {Kolotkov}, {Nakariakov},
		{Molevich}  \& {Ryashchikov}}{{Zavershinskii}
		et~al.}{2019}]{2019PhPl...26h2113Z}
	{Zavershinskii} D.~I.,  {Kolotkov} D.~Y.,  {Nakariakov} V.~M.,  {Molevich}
	N.~E.,   {Ryashchikov} D.~S.,  2019, \mn@doi [Physics of Plasmas]
	{10.1063/1.5115224}, \href
	{https://ui.adsabs.harvard.edu/abs/2019PhPl...26h2113Z} {26, 082113}
	
	\bibitem[\protect\citeauthoryear{{Zavershinskii}, {Kolotkov}, {Riashchikov}  \&
		{Molevich}}{{Zavershinskii} et~al.}{2021}]{2021SoPh..296...96Z}
	{Zavershinskii} D.,  {Kolotkov} D.,  {Riashchikov} D.,   {Molevich} N.,  2021,
	\mn@doi [\solphys] {10.1007/s11207-021-01841-1}, \href
	{https://ui.adsabs.harvard.edu/abs/2021SoPh..296...96Z} {296, 96}
	
	\bibitem[\protect\citeauthoryear{{Zhang}, {Zhang}, {Wang}  \&
		{Nakariakov}}{{Zhang} et~al.}{2015}]{2015A&A...581A..78Z}
	{Zhang} Y.,  {Zhang} J.,  {Wang} J.,   {Nakariakov} V.~M.,  2015, \mn@doi
	[\aap] {10.1051/0004-6361/201525621}, \href
	{https://ui.adsabs.harvard.edu/abs/2015A&A...581A..78Z} {581, A78}
	
	\bibitem[\protect\citeauthoryear{{Zimovets} et~al.,}{{Zimovets}
		et~al.}{2021}]{2021SSRv..217...66Z}
	{Zimovets} I.~V.,  et~al., 2021, \mn@doi [\ssr] {10.1007/s11214-021-00840-9},
	\href {https://ui.adsabs.harvard.edu/abs/2021SSRv..217...66Z} {217, 66}
	
	\makeatother
\end{thebibliography}

\bsp	
\label{lastpage}
\end{document}